\documentclass[proceedings]{JHEP37}
\usepackage{eurosym}
\usepackage{amssymb}
\usepackage{amsfonts,cite}
\usepackage{amsmath}
\usepackage{graphicx}
\usepackage{multirow}
\usepackage{epsfig}
\DeclareMathOperator\arccosh{arccosh}
\DeclareMathOperator\sech{sech}

\newbox\mybox

\newcommand\fverb{\setbox\mybox=\hbox\bgroup\verb}
\newcommand\fverbdo{\egroup\medskip\noindent\fbox{\unhbox\mybox}\ }
\newcommand\fverbit{\egroup\item[\fbox{\unhbox\mybox}]}
\conference{Stability of complex soliton solutions in the Bullough-Dodd model}
\abstract{We investigate different types of complex soliton solutions with regard to their stability against linear pertubations. In the Bullough-Dodd scalar field theory we find linearly stable complex ${\cal{PT}}$-symmetric solutions and linearly unstable solutions for which the ${\cal{PT}}$-symmetry is broken. Both types of solutions have real energies. 
	The auxiliary Sturm-Liouville eigenvalue equation in the stability analysis for the ${\cal{PT}}$-symmetric solutions can be solved exactly by supersymmetrically mapping it to an isospectral partner system involving a shifted and scaled inverse $\cosh$-squared potential.
	We identify exactly one shape mode in form of a bound state solution and scattering states which when used as linear perturbations leave the solutions stable. The auxiliary problem for the solutions with broken ${\cal{PT}}$-symmetry involves a complex shifted and scaled inverse $\sin$-squared potential. The corresponding bound and scattering state solutions have complex eigenvalues, such that when used as linear perturbations for the corresponding soliton solutions lead to their decay or blow up as time evolves.}

\title{Linearly stable and unstable complex soliton solutions with real energies in the Bullough-Dodd model}
\author{Francisco Correa$^\circ$, Andreas Fring$^\bullet$ and Takanobu Taira$%
^\bullet$ \\
$\bullet$ Department of Mathematics, City, University of London,\\
$\,\,$ Northampton Square, London EC1V 0HB, UK \\
$\circ$ Instituto de Ciencias F{\'{\i}}sicas y Matem{\'{a}}ticas,
Universidad Austral de Chile, \\
$\,\,$ Casilla 567, Valdivia, Chile\\
E-mail: francisco.correa@uach.cl, a.fring@city.ac.uk,
takanobu.taira@city.ac.uk}

\input{tcilatex}
\begin{document}

\section{Introduction}
The success of ${\cal{PT}}$-symmetric quantum mechanics \cite{PTbook} and the applications of the concepts in optical analogues \cite{el2018non} have inspired research in many different areas of physics. One of them is the study of complex ${\cal{PT}}$-symmetric soliton solutions in nonlinear integrable systems in 1+1 dimensions \cite{AFKdV,wadati2008construction,CompPaulo,CFB,CenFring,correa2016regularized,khare2016novel,ablowitz2013,CenFringHir,JuliaThesis,fring2020BPS} or models that admit complex Skyrmion type solutions in 2+1 dimensions \cite{CFTSkyrmions}. Despite being complex, these sort of solutions possess nonetheless real energies, a property that can be attributed to general arguments that make use of the underlying antilinear symmetry, generally referred to as ${\cal{PT}}$-symmetry \cite{AFKdV,CenFring,fring2020BPS,CFTSkyrmions}.
 
The other aspect, that makes such type of solutions potentially interesting physical objects, is their stability. So far this vital property has been largely ignored in their studies and here the main purpose is to investigate the stability properties of particular complex one-soliton solutions for a specific integrable scalar field theory, the Bullough-Dodd model \cite{BDodd,zhiber1979klein}. Besides shedding light on the important question of whether a complex soliton solution will collapse, i.e. grow or decay, when slightly perturbed, it is also well-known that as a by-product of a stability analysis one may obtain the so-called shape mode as a bound state solution for the auxiliary Sturm-Liouville eigenvalue equation. These solutions are instrumental in the understanding of energy transfer effects in the dynamics of multi-soliton solutions that can be unravelled using a collective coordinate or moduli space approach  \cite{takyi2016coll,manton2021kink}. For the model studied here we will investigate these moduli space properties in upcoming work \cite{CFTprep}. 
  
Our manuscript is organised as follows: In section 2 we assemble some generalities about the stability analysis to be carried out. In addition, we briefly recall the general set up for a supersymmetric formulation of quantum mechanics. While this is not essential for a stability analysis, it greatly facilitates the construction of exact solutions to the auxiliary Sturm-Liouville eigenvalue equations. In section 3 we construct complex one-soliton solutions for the Bullough-Dodd model and analyse in detail their stability.

\section{Stability analysis, zero modes, shape modes and supersymmetry}

The systems considered here are complex scalar field theories with a Lagrangian density of the
general form%
\begin{equation}
\mathcal{L=}\frac{1}{2}\partial _{\mu }\varphi \partial ^{\mu }\varphi
-V(\varphi ),  \label{lag}
\end{equation}%
with $\varphi (x,t)\in \mathbb{C}$ and potential $V(\varphi )\in \mathbb{C}$. We denote the Lagrangian, i.e. the volume integral over this density, as $%
L=\int\nolimits_{-\infty }^{\infty }$ $\mathcal{L}dx$. Our space-time metric
is taken to be Lorentzian of the form $\limfunc{diag}(1,-1)$, so that the
Euler-Lagrange equation resulting from (\ref{lag}) reads 
\begin{equation}
\ddot{\varphi}-\varphi ^{\prime \prime }+\frac{\partial V(\varphi )}{%
\partial \varphi }=0.  \label{EL}
\end{equation}%
We use standard conventions and denote partial derivatives with respect to
time $t$ and space $x$ by overdots and dashes, respectively. The static
solutions to these equations, i.e. the solutions to (\ref{EL}) with $\dot{%
\varphi}=0$, are fields $\phi (x,\mathbf{u})$ depending on the space coordinate $x$ and a
set of constants that are identified as moduli or collective coordinates $%
\mathbf{u}=(u^{1},\ldots ,u^{n})$. As is well-known, one may construct the time-dependent
solutions $\varphi $ from the time-independent solutions $\phi $ simply by a
Lorentz boost $\phi \lbrack (x-vt)/\sqrt{1-v^{2}}]=\varphi (x,t)$ with $v$
denoting the velocity. The energies of particular solutions are computed as
\begin{equation}
	E[\varphi]= \int\nolimits_{-\infty }^{\infty } dx  \varepsilon(\varphi),   \qquad  \varepsilon(\varphi)= \left( \frac{1}{2}\dot{\varphi}^{2}  +\frac{1}{2}  (\varphi^{\prime } )^2  +V(\varphi )  \right),   \label{ener}
\end{equation}%
where $\varepsilon(\varphi)$ denotes the energy density. When the Hamiltonian associated to (\ref{lag}) admits a modified ${\cal{CPT}}$-symmetry and $\varphi$ is mapped by this symmetry to itself or possibly to a second degenerate solution, the energy $E[\varphi]$ is guaranteed to be real, irrespective of whether $\varphi \in \mathbb{R}$, $\varphi \in \mathbb{C}$ or $V^\dagger(\varphi) = V(\varphi)$ \cite{CFTSkyrmions}. This reasoning may also be extended to ensure the reality of other conserved quantities.

With regard to the main purpose of this paper, we now briefly recall the main argument of a standard stability analysis
beginning with the linearisation of the Euler-Lagrange equation (\ref{EL})
by replacing 
\begin{equation}
\varphi \rightarrow \varphi _{s}+\varepsilon \delta \varphi \qquad
\varepsilon \ll 1,  \label{lin}
\end{equation}%
where $\varphi _{s}$ solves (\ref{EL}) and $\delta \varphi =:\chi $ is a
small perturbation. This converts the Euler-Lagrange equation into 
\begin{equation}
\ddot{\varphi _{s}}-\varphi _{s}^{\prime \prime }+\left. \frac{\partial
V(\varphi )}{\partial \varphi }\right\vert _{\varphi _{s}}+\varepsilon
\left( \ddot{\chi}-\chi ^{\prime \prime }+\left. \chi \frac{\partial
^{2}V(\varphi )}{\partial \varphi ^{2}}\right\vert _{\varphi _{s}}\right) +%
\mathcal{O}(\varepsilon ^{2})=0.  \label{ELeps}
\end{equation}%
Evidently the first three terms in (\ref{ELeps}) vanish by construction and
when assuming the perturbation to be separable of the form $\chi
(x,t)=e^{i\lambda t}\Phi (x)$, the term of first order in $\varepsilon $
reduces to a Sturm-Liouville eigenvalue problem of the same form as the
time-independent Schr\"{o}dinger equation 
\begin{equation}
-\Phi _{xx}+V_{1}\Phi =\lambda ^{2}\Phi ,~~~~\ \ \ \ \ \text{with \ }%
V_{1}(x):=\left. \frac{\partial ^{2}V(\varphi )}{\partial \varphi ^{2}}%
\right\vert _{\varphi _{s}}.  \label{tisch}
\end{equation}%
For the eigenfunction solutions to this equation with $\lambda \in \mathbb{R}$
the linear perturbation will simply introduce an oscillation in time around the
solution $\varphi _{s}$, whereas when $\lambda \in \mathbb{C}$ the solutions
will grow or decay with time and are therefore unstable. 

Let us next see how the energy of the perturbed solution behaves. Expanding up to second order in $\varepsilon$ and integrating two terms by parts, we easily derive
\begin{eqnarray}
		E[\varphi_s + \chi] &=& 	E[\varphi_s ] + \!\!\! \int\limits_{-\infty }^{\infty } dx   \left[ \left(  \left. \frac{\partial
			V(\varphi )} {\partial \varphi }\right\vert _{\varphi _{s}} - \varphi_s^{\prime \prime }  \right) \chi 
		+ \frac{1}{2} \chi \left( \frac{\dot \chi ^2}{\chi} - \chi^{\prime \prime } +  \left. \chi \frac{\partial
			^{2}V(\varphi )}{\partial \varphi ^{2}}\right\vert _{\varphi _{s}}  \right) \right] \,\,\,\,\, \, \,\, \label{enn} \\
		&& + \left. \chi ( \chi^{\prime} + \varphi_s^{\prime})     \right\vert_{- \infty}^{\infty} + \mathcal{O}(\varepsilon ^{3}) . \notag
\end{eqnarray}
We notice that first bracket of the integrand is simply the Euler-Lagrange equation for the static solution. The second bracket of the integrand vanishes when $\chi$ is separable as stated above and $\Phi (x)$ satisfies the Sturm-Liouville equation (\ref{tisch}). The surface term in the second line of (\ref{enn}) is zero when $\Phi (x)$ vanishes asymptotically $\lim_{x \rightarrow \pm \infty} \Phi (x) = 0$, so that the energy of the perturbed solution becomes identical to the one of the static solution $\varphi_s$. 

The \emph{zero mode}, or translation mode, defined as the solution $\Phi _{0}$ to (\ref{tisch}) with $\lambda =0$, plays
a special role for a number of reasons. First of all, it is well-known \cite%
{jackiw1977quantum,weinberg1979parameter,kim1990index} that the
dimensionality of the corresponding moduli spaces may be obtained from these
modes. Moreover it allows to directly compute the superpotential and to set
up a supersymmetric scheme that will be important for the concrete system we
consider below.

The zero mode is easily computed: Assuming that the static solution $\phi (x)$ belongs to a class of solutions
characterised by a continuous parameter, say $a$, such that $\phi (x)=\psi
(x,a)|_{a=0}$, we may think of $\phi _{s}+\varepsilon \delta \phi $ in (\ref%
{lin}) as 
\begin{equation}
\psi (x,a)=\phi (x)+a\left. \frac{\partial \psi (x,a)}{\partial a}%
\right\vert _{a=0}+\mathcal{O}(a^{2}),  \label{stzero}
\end{equation}%
so that we can identify the zero mode as 
\begin{equation}
\Phi _{0}(x):=\left. \frac{\partial \psi (x,a)}{\partial a}\right\vert
_{a=0}.  \label{zeromodecal}
\end{equation}%
We may now repeat the argument for different continuous parameters $%
a\rightarrow a_{1},\ldots a_{n}$, so that the number $n$ of independent zero
modes constitutes the dimension of the moduli space.

The excited bound states or scattering states for the eigenvalue equation in (\ref{tisch}) may be constructed in a number of ways. Here it turns out that the potentials we are interested in possess superpartner potentials that have been solved in the literature, so that we may employ the scheme and simply map the known solutions to our problem. Let us therefore briefly recall the
general set up of supersymmetric quantum mechanics by following the ideas of Witten and others \cite%
{Witten:1981nf,Cooper:1982dm,Witten,Cooper}.  Assuming that the eigenvalue equation (\ref%
{tisch}) has a discrete eigenvalue spectrum, $E_{n}^{(1)}=\lambda _{n}^{2}$,
we re-write it as%
\begin{equation}
H_{1}\Phi _{n}^{(1)}=\left( -\frac{d^{2}}{dx^{2}}+V_{1}\right) \Phi
_{n}^{(1)}=L_{+}L_{-}\Phi _{n}^{(1)}=E_{n}^{(1)}\Phi _{n}^{(1)},  \label{H1}
\end{equation}%
with $\Phi _{0}=:\Phi _{0}^{(1)}$ and $E_{0}=0$. We assumed here that the
Hamiltonian factorises into the product of two first order
differential operators defined as
\begin{equation}
L_{\pm }:=\mp \frac{d}{dx}+W(x),~~\ \ \ \text{with \ }W(x):=-\frac{\Phi
_{0}^{\prime }}{\Phi _{0}}, \label{LW}
\end{equation}%
where $W(x)$ is referred to as the {\em superpotential} related to the potential
in (\ref{H1}) as $V_{1}=W^{2}-W^{\prime }$. Defining next a new Hamiltonian $H_2$
with the product of the operators $L_{\pm }$ in reverse order, we have%
\begin{equation}
H_{2}\Phi _{m}^{(2)}=\left( -\frac{d^{2}}{dx^{2}}+V_{2}\right) \Phi
_{m}^{(2)}=L_{-}L_{+}\Phi _{m}^{(2)}=E_{m}^{(2)}\Phi _{m}^{(2)},  \label{H2}
\end{equation}%
with a second potential  $V_{2}=W^{2}+W^{\prime }$. Considering now 
\begin{eqnarray}
H_{2}\left( L_{-}\Phi _{n}^{(1)}\right)  &=&L_{-}L_{+}L_{-}\Phi
_{n}^{(1)}=E_{n}^{(1)}\left( L_{-}\Phi _{n}^{(1)}\right) , \\
H_{1}\left( L_{+}\Phi _{m}^{(2)}\right)  &=&L_{+}L_{-}L_{+}\Phi
_{m}^{(2)}=E_{m}^{(2)}\left( L_{+}\Phi _{m}^{(2)}\right) ,
\end{eqnarray}%
and comparing these equations with (\ref{H1}), (\ref{H2}) we conclude%
\begin{equation}
\Phi _{n+1}^{(1)}=N_{n+1}^{(1)}L_{+}\Phi _{n}^{(2)},~~\ ~\Phi
_{n}^{(2)}=N_{n}^{(2)}L_{-}\Phi _{n}^{(1)},~~\
~E_{n}^{(2)}=E_{n+1}^{(1)},~~\ ~E_{0}^{(1)}=0,~~~n\in \mathbb{N}_{0}.
\label{Super}
\end{equation}%
Thus apart from $E_{0}^{(1)}$ the two Hamiltonians $H_{1}$ and $H_{2}$ are
isospectral, hence the referral as {\em supersymmetric}. When $L_{-}^{\dagger
}=L_{+}$ the normalisation constants are simply $N_{n+1}^{(1)}=1/\sqrt{%
E_{n}^{(2)}}$, $N_{n}^{(2)}=1/\sqrt{E_{n+1}^{(1)}}$, but we will not assume
this as we allow here for complex superpotentials $W$. The scheme is directly extended to scattering solutions with continuous eigenvalues spectra.
 
The crucial feature for
our purposes is here the observation from (\ref{Super}) that when
one of the eigenvalue problems (\ref{H1}) or (\ref{H2}) has been solved, one
may directly deduce the solutions for the other. It is in this sense that we
will apply this scheme as the potentials $V_{1}$ we construct from the zero
modes are not obvious to solve, whereas the solutions for the supersymmetric partner
potentials $V_{2}$ can be found in the literature.

\section{The Bullough-Dodd model}

As an example, we are interested here in the stability properties of the complex soliton solutions of the Bullough-Dodd model \cite{BDodd,zhiber1979klein}, that
is a one scalar field integrable field theory already quite well studied in its
classical \cite{andreevback} and quantum \cite{Fring:1992pj} aspects. It is
described by the Lagrangian density of the form
\begin{equation}
\mathcal{L}_{\text{BD}} =\frac{1}{2}\partial _{\mu }\varphi \partial ^{\mu }\varphi
-e^{\varphi }-\frac{1}{2}e^{-2\varphi }+\frac{3}{2} \qquad \text{with} \;\; \varphi \in {\mathbb{C}} . \label{BDLag}
\end{equation}%
The resulting classical nonlinear equation of motion 
\begin{equation}
\ddot{\varphi}-\varphi ^{\prime \prime }+e^{\varphi }-e^{-2\varphi }=0,
\label{BDequ}
\end{equation}%
can be solved by standard techniques from classical theory of integrable systems.
Here we exploit the fact that the equation may be solved using Hirota's
direct method \cite{hirota2004direct} when parameterizing the fields as $%
\varphi =\ln (\tau _{0}/\tau _{1})$ with $\tau _{i}$ denoting the so-called $%
\tau $-functions. To identify various one-soliton solutions we can therefore
make the general Ansatz 
\begin{equation}
\varphi (x,t)=\ln \left( \frac{\alpha _{0}+\alpha _{1}e^{kx+lt}+\alpha
_{2}e^{2kx+2lt}}{\beta _{0}+\beta _{1}e^{kx+lt}+\beta _{2}e^{2kx+2lt}}%
\right) ,  \label{Ansatz}
\end{equation}%
with unknown constants $\alpha _{i},\beta _{i}$, $i=0,1,2$, and $k,l$.
Substituting this Ansatz into equation (\ref{BDequ}) and subsequently
reading off the coefficients of $e^{jkx+jlt}$ for $j=0,\ldots ,8$ leads to 9
equations, that we do not report here, which may solved for the unknown
constants in (\ref{Ansatz}). In this manner we find various types of solutions.

\subsection{Complex one-soliton solutions of type I} \label{onesolrealpure}
\begin{figure}[h]
	\centering         
	\begin{minipage}[b]{0.52\textwidth}           \includegraphics[width=\textwidth]{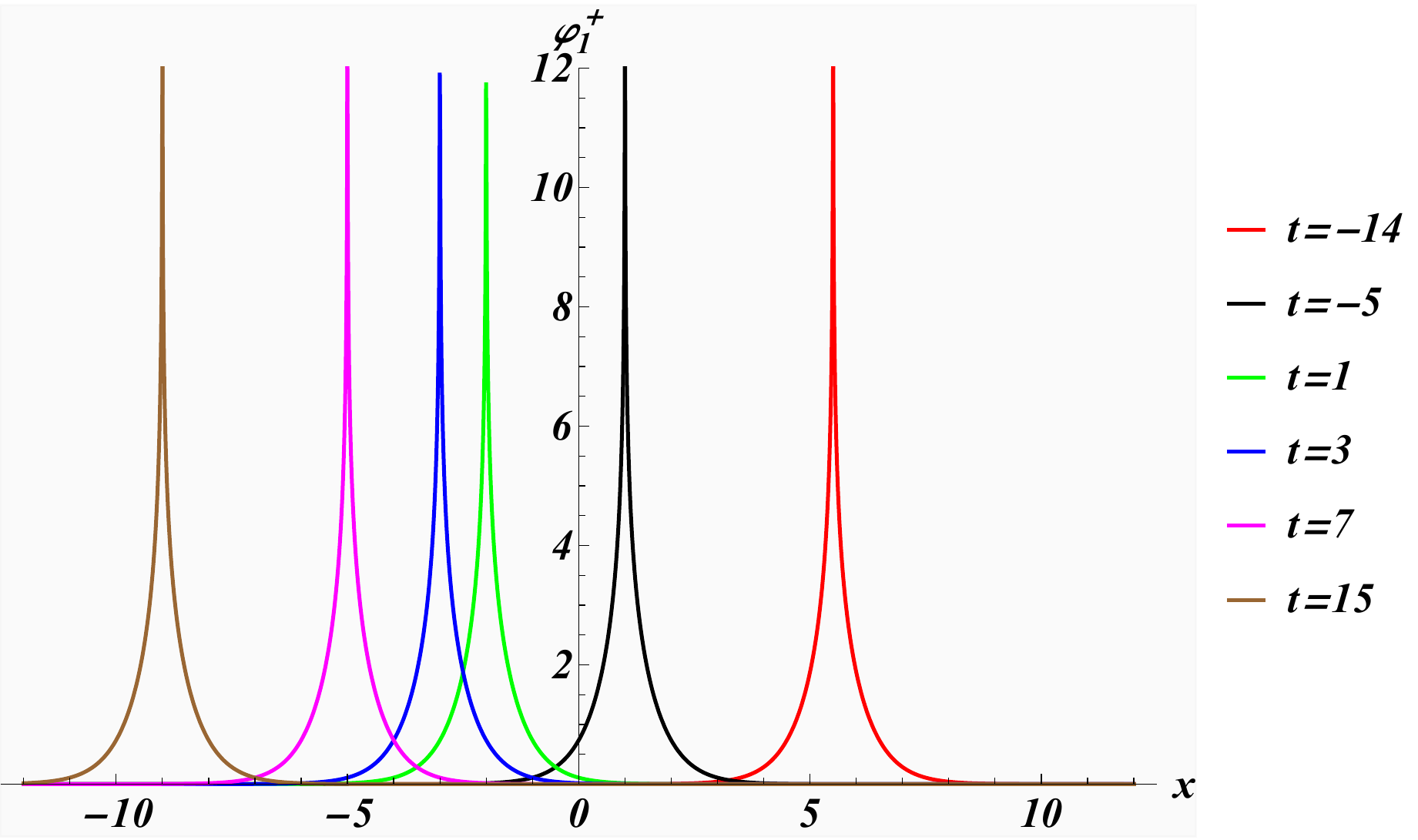}
	\end{minipage}   
	\begin{minipage}[b]{0.44\textwidth}           
		\includegraphics[width=\textwidth]{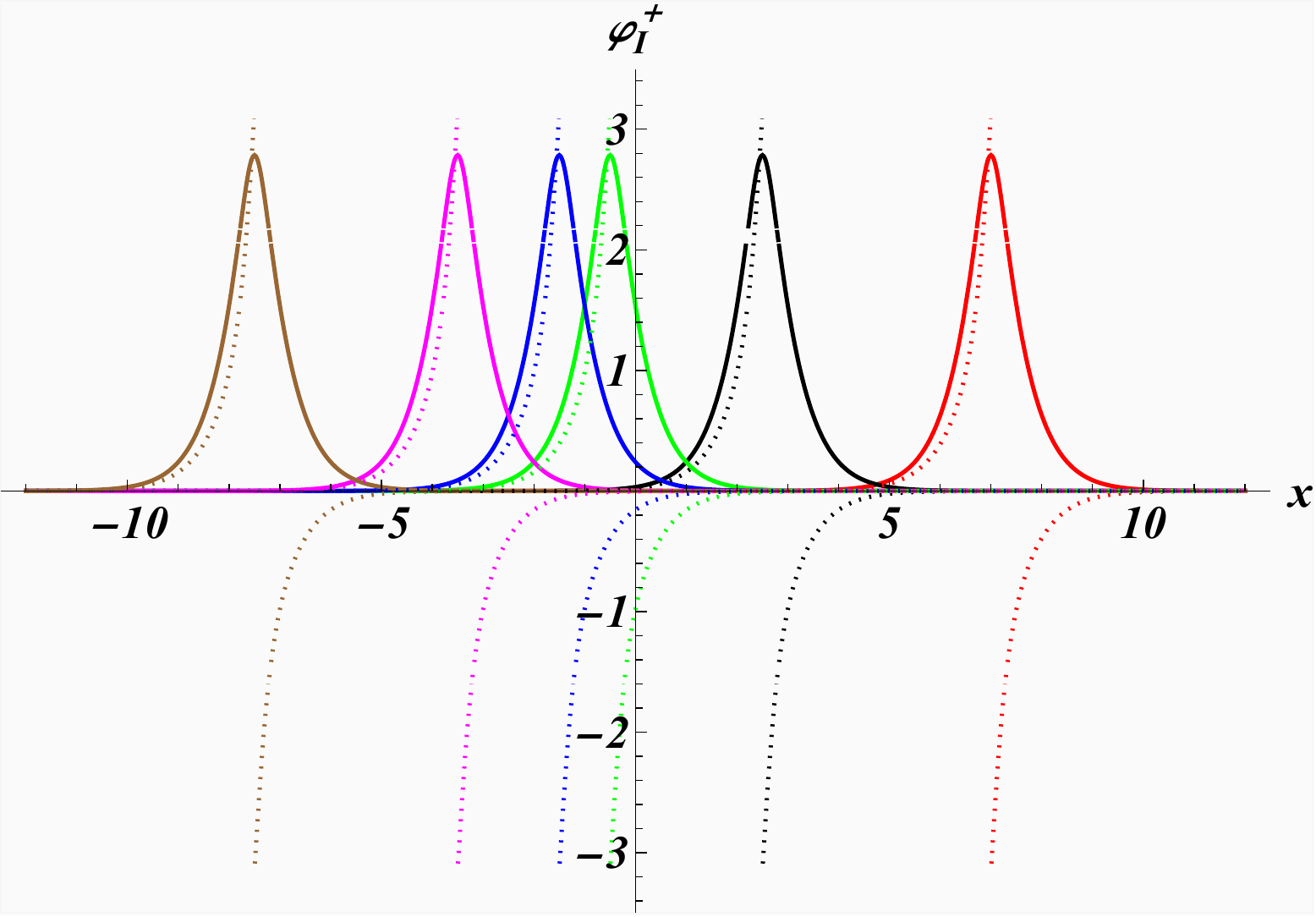}
	\end{minipage}    
	\caption{Scattering solutions (\ref{sol1}) of the Bullough-Dodd classical equation of motion (\ref{BDequ}) for $|k|> \sqrt{3}$. Panel (a): Travelling real one-cusp solution $\varphi_I^{+}$ for $k=2$ with $\beta=3$ and panel (b) ${\cal{PT}}$ regularized travelling complex solution $\varphi_I^{+}$ for $k=2$ with $\beta=i 3/5$. Solid lines correspond to real and dotted lines to imaginary parts. }
	\label{BDsolution1}
\end{figure}

\begin{figure}[h]
	\centering         
	\begin{minipage}[b]{0.52\textwidth}           \includegraphics[width=\textwidth]{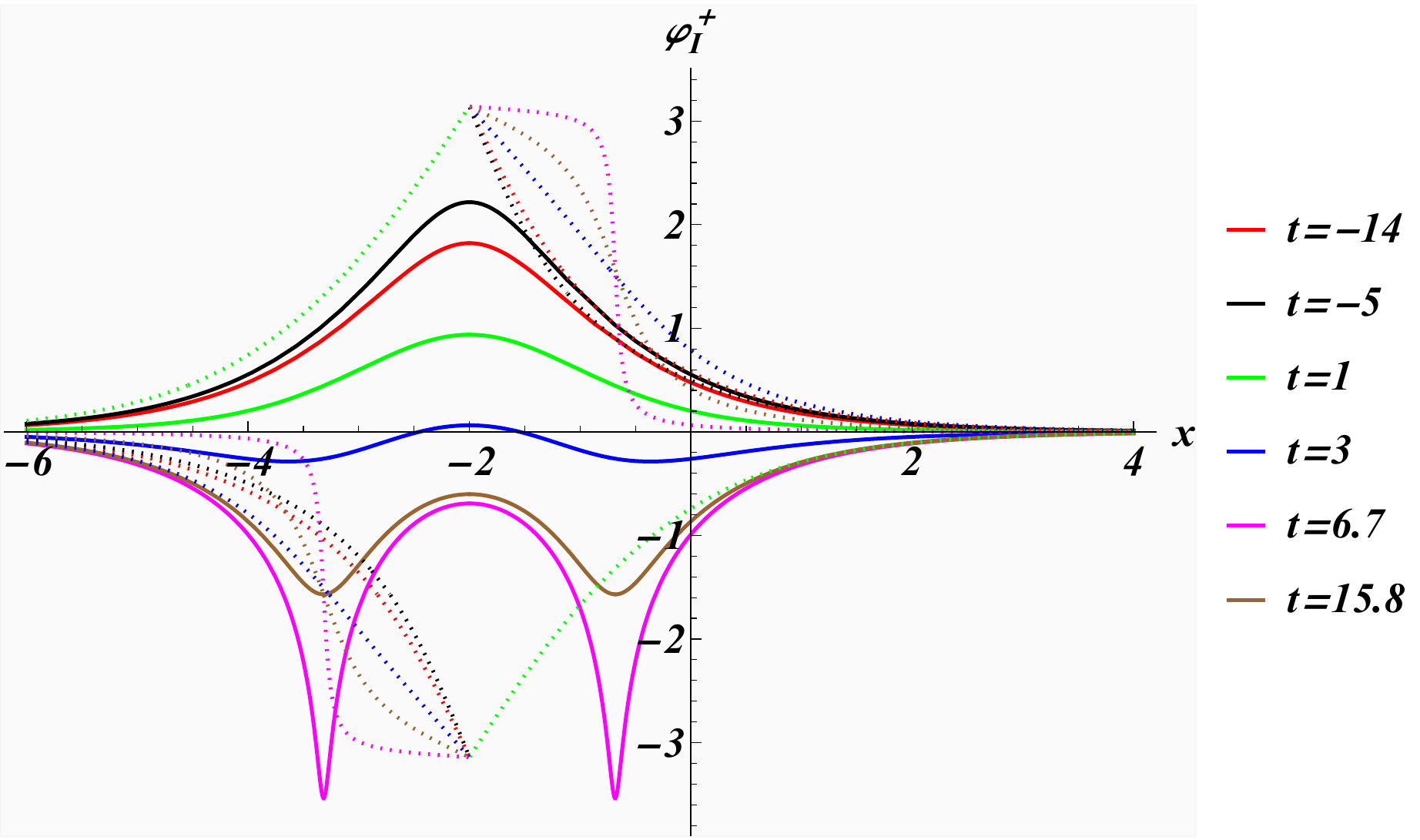}
	\end{minipage}   
	\begin{minipage}[b]{0.44\textwidth}           
		\includegraphics[width=\textwidth]{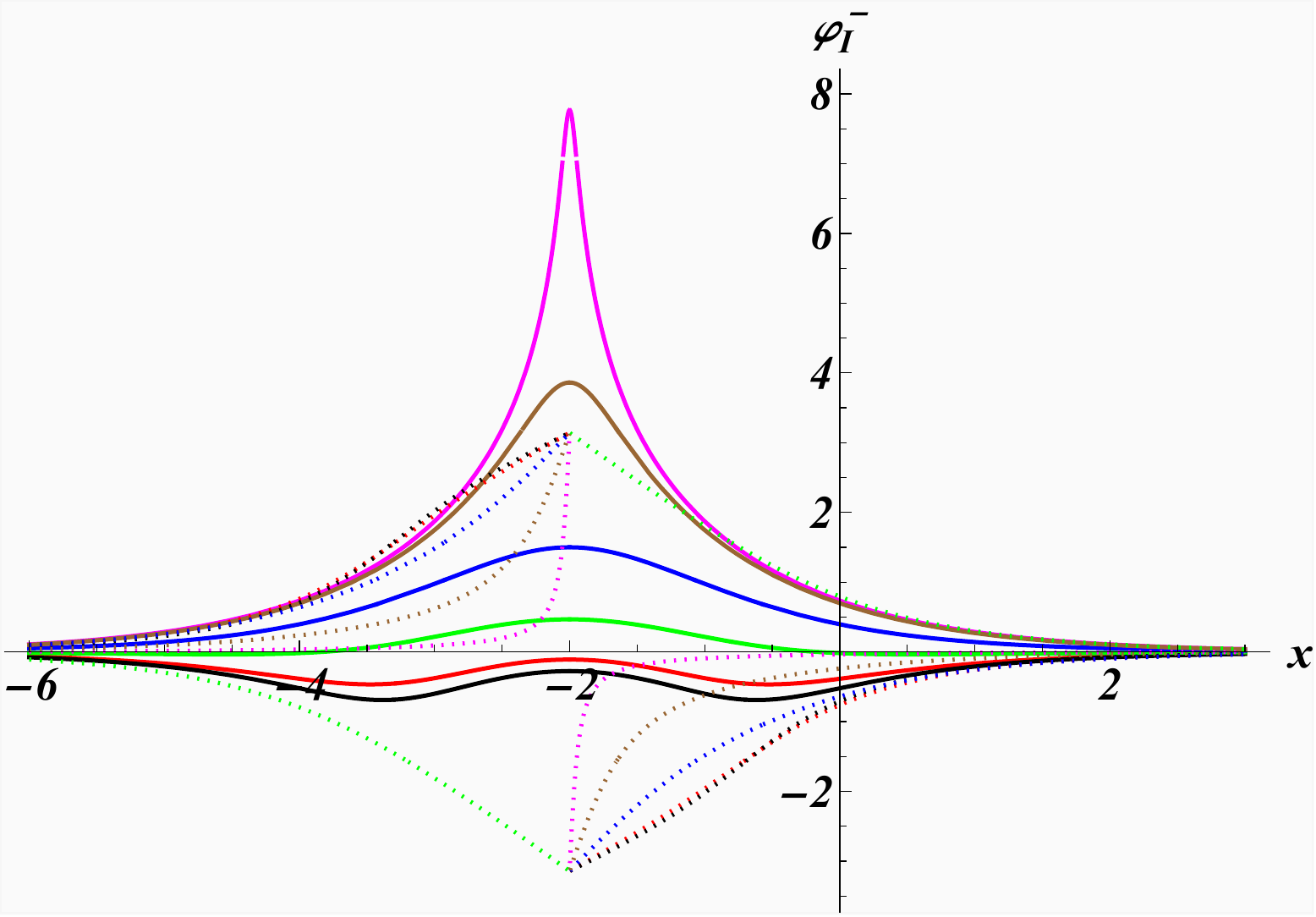}
	\end{minipage}    
	\caption{Complex breather solutions (\ref{sol1}) of the Bullough-Dodd classical equation of motion (\ref{BDequ}) for $|k|< \sqrt{3}$. Panel (a): $\varphi_I^{+}$ for $k=1< \sqrt{3}$ with $\beta=2$ and panel (b) $\varphi_I^{-}$ for $k=1< \sqrt{3}$ with $\beta=2$. Solid lines correspond to real and dotted lines to imaginary parts.}
	\label{BDsolution2}
\end{figure}

The first type of solutions we obtain are one-soliton solutions of the form
\begin{equation}
\varphi_I^{\pm} (x,t)= \ln \left[\frac{\cosh \left(\beta +\sqrt{k^2-3} t+k x\right) \pm 2}{\cosh \left(\beta +\sqrt{k^2-3} t+k
	x\right)\mp 1}\right],  \quad \beta \in  \mathbb{C}. \label{sol1}
\end{equation}%
When $\beta$ is real and $|k|> \sqrt{3}$ the solution $\varphi_I^{+}$ is real but becomes singular for
$x_0=-[ \beta +\sqrt{k^2-3} t]/k$, whereas $\varphi_I^{-}$ is only real for $x< x_l= -[ \beta +\arccosh(2) +\sqrt{k^2-3} t]/k$ and $x> x_r=[-\beta +\arccosh(2) -\sqrt{k^2-3} t]/k$ as the argument of the logarithm becomes negative in the complementary regime. Samples of these types of solutions in the different regimes are depicted in figure \ref{BDsolution1}. In the case $|k|< \sqrt{3}$ both solutions $\varphi_I^{\pm}$ are always complex, see figure \ref{BDsolution2}. The type I solutions found here formally coincide with the solutions constructed in \cite{assis2008bullough}.

\begin{figure}[h]
	\centering         
	\begin{minipage}[b]{0.52\textwidth}           \includegraphics[width=\textwidth]{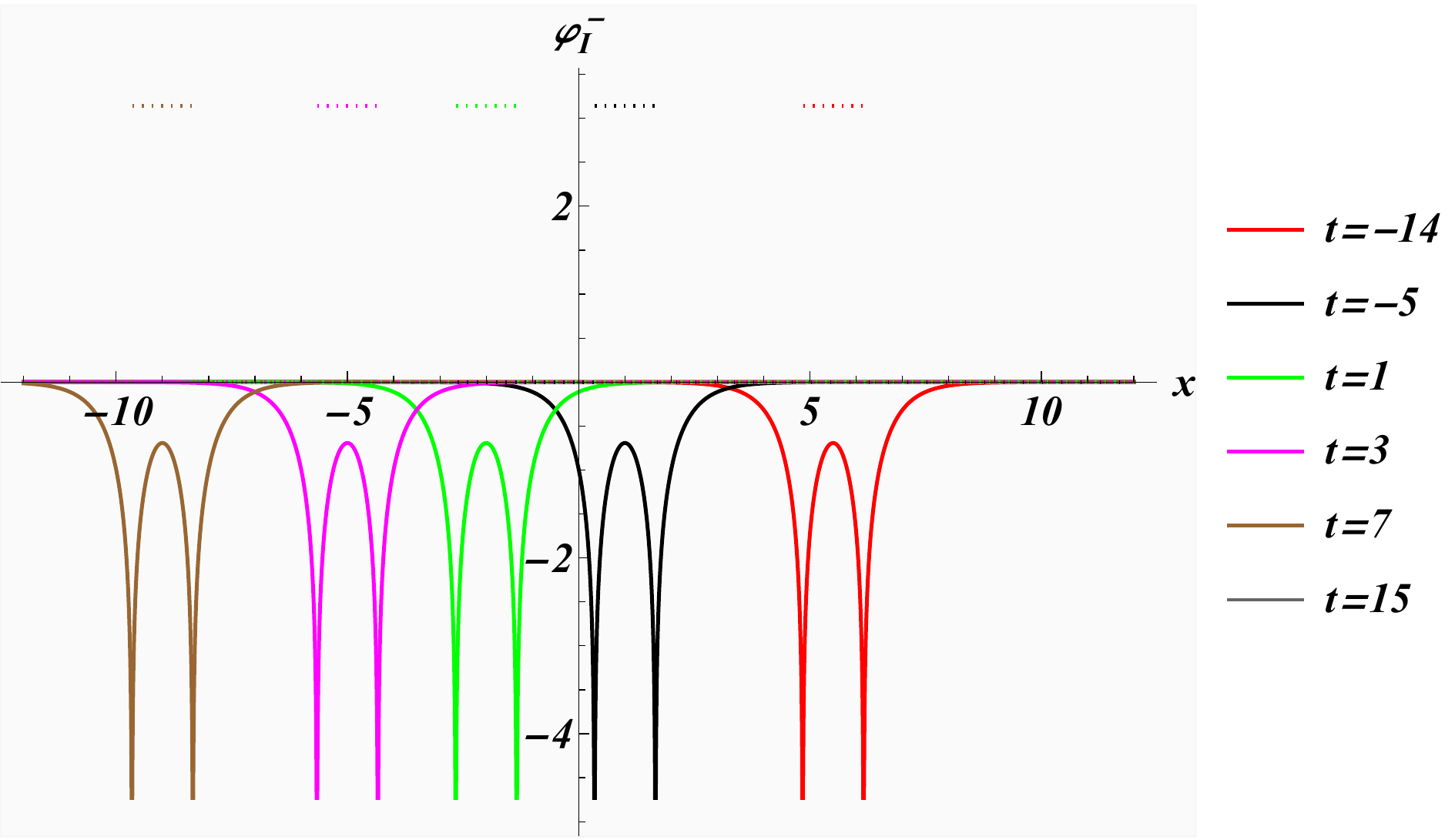}
	\end{minipage}   
	\begin{minipage}[b]{0.44\textwidth}           
		\includegraphics[width=\textwidth]{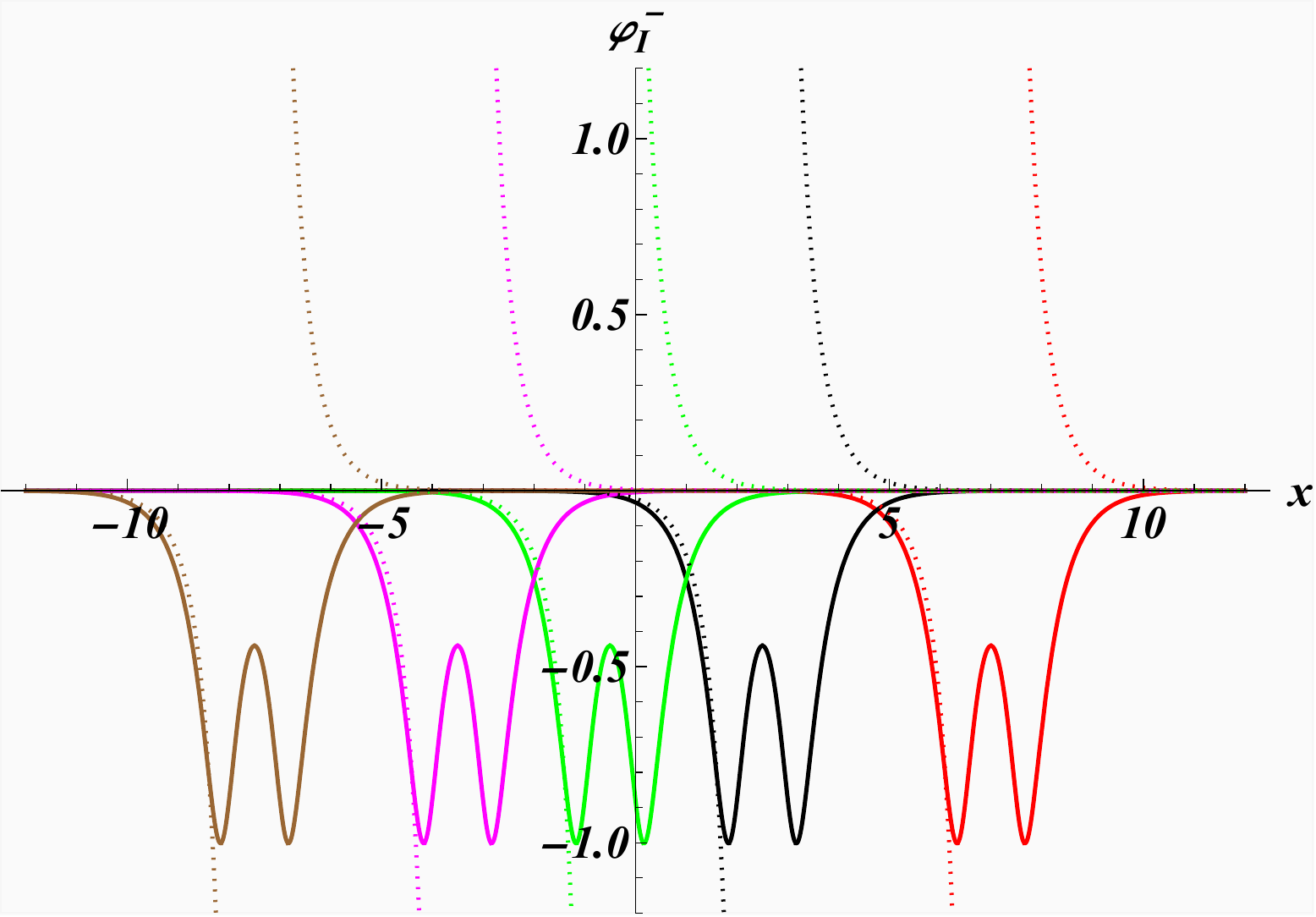}
	\end{minipage}    
	\caption{Scattering solutions (\ref{sol1}) of the Bullough-Dodd classical equation of motion (\ref{BDequ}) for $|k|> \sqrt{3}$. Panel (a): Complex solution $\varphi_I^{-}$ for $k=2$ with $\beta=3$ and panel (b) ${\cal{PT}}$ regularized complex solution $\varphi_I^{-}$ for $k=2$ with $\beta=i 3/5$. Solid lines correspond to real and dotted lines to imaginary parts. }
	\label{BDsolution1b}
\end{figure}

Despite the fact that these solutions are complex, their energies are real governed by an antilinear ({$\cal{PT}$}-symmetry), and for N-soliton solutions together with the fact that these theories are integrable \cite{CenFring,fring2020BPS}. The {$\cal{PT}$}-symmetry is easily identified as 
\begin{equation}
	{\cal{PT}}: x \rightarrow -x, \quad t \rightarrow -t, \quad i \rightarrow -i, \quad \varphi \rightarrow \varphi . \label{PT}
\end{equation}
The Bullough-Dodd Lagrangian (\ref{BDLag}) is trivially invariant under this symmetry and the solutions $\varphi_I^{\pm} (x,t)$ evidently respect it for purely imaginary constants $\beta$, i.e. ${\cal{PT}}: \varphi_I^{\pm} \rightarrow \varphi_I^{\pm}$. Thus to take $\beta \in i \mathbb{R}$ seems to be very suggestive as this choice not only converts the solutions into ${\cal{PT}}$-symmetric ones, but also regularises the singularities of the solutions at the cost of a discontinuity resulting from the branch cut of the logarithm as depicted in figures \ref{BDsolution1} and \ref{BDsolution1b}. The behaviour of these complex shifted solutions is identical to those for the $|k|< \sqrt{3}$ and $\beta \in  \mathbb{R}$ already depicted in figure \ref{BDsolution2}, when we identify in these solutions $\beta \rightarrow \sqrt{k^2-3}t $ and $\sqrt{k^2-3}t \rightarrow i \beta$ for fixed time.

Indeed the energies for the solutions (\ref{sol1}), evaluated according to (\ref{ener}), are always real
\begin{equation}
 E [ \varphi_I^{\pm} ] = - 6 |k| , \label{energ}
\end{equation}
and independent of the complex parameter $\beta$.

\subsubsection{Zero modes for type I complex one-soliton solutions}

The Sturm-Liouville eigenvalue problem (\ref{tisch}) of the stability analysis for the Bullough-Dodd model reads 
\begin{equation}
	- \Phi_{xx} +  V(x) \Phi = \lambda^2 \Phi \qquad \text{with} \quad V(x)=\left( e^{\phi(x)} + 2 e^{-2 \phi(x)}  \right) . \label{BDtisch}
\end{equation}
The static solution for the cusp and oscillatory solutions $\varphi^{\pm}_I(x,t)$ acquire the form
\begin{equation}
	\phi^{\pm}_I(x) = \ln \left[\frac{\cosh \left(\beta +\sqrt{3} x\right) \pm 2}{\cosh \left(\beta +\sqrt{3} x\right) \mp 1}\right], 
\end{equation}
which when substituted into (\ref{BDtisch}) yield the potentials
\begin{eqnarray}
	V_1^{+}(x) &=& 1-\frac{3}{1 - \cosh \left(\beta +\sqrt{3} x\right)}+\frac{8 \sinh ^4\left[\frac{1}{2} \left(\beta +\sqrt{3}
		x\right)\right]}{\left[2 + \cosh \left(\beta +\sqrt{3} x\right) \right]^2}, \\
		V_1^{-}(x) &=& 1-\frac{3}{1 + \cosh \left(\beta +\sqrt{3} x\right)}+\frac{8 \cosh ^4\left[\frac{1}{2} \left(\beta +\sqrt{3}
		x\right)\right]}{\left[2 - \cosh \left(\beta +\sqrt{3} x\right) \right]^2}.  \label{pot}
\end{eqnarray}
We think now of the larger class of functions depending on the continuous parameter $a$ as $\psi^{\pm}(x,a)=\phi^{\pm}(x+a/\sqrt{3})$. By means of (\ref{zeromodecal}) the zero mode is then easily computed to
\begin{equation}
	\Phi_0^{\pm}(x) = - \frac{3 \left\{ \tanh \left[\frac{1}{2} \left(\beta +\sqrt{3} x\right)\right] \right\}^{\mp 1} }{2 \pm \cosh \left(\beta +\sqrt{3} x\right) } . \label{phizero}
\end{equation}
One verifies that the $\Phi_0^{\pm}(x)$ indeed satisfy (\ref{BDtisch}) for the potentials (\ref{pot}) with eigenvalue $\lambda=0$. Notice that when taking $a \rightarrow i a$, as seems to be natural when $k < \sqrt{3}$, so that we identify  $\psi^{\pm}(x,a)=\phi^{\pm}(x+i a/\sqrt{3})$, the resulting zero mode is the same as in (\ref{phizero}) only multiplied by $i$, which simply corresponds to a different normalization constant. 

\subsubsection{Bound state solutions for type I complex one-soliton solutions}
It is not straightforward or obvious how to solve the eigenvalue equation for the potentials $V_1^{\pm}(x)$ directly. However, we notice now that the supersymmetric partner Hamiltonians are inverse $\cosh^2$-potentials that have been treated extensively in the literature. Since $V_1^{+}(x, \beta)=V_1^{-}(x, \beta + i \pi)$, we just focus on the ``+"-case from now on, as all quantities related to the ``-"-system are computed simply by a shift in $\beta$. Using the relation (\ref{LW}) we compute the superpotential from the zero modes to
\begin{equation}
W^{+}(x) =  \frac{\sqrt{3}}{2} \frac{ 3+ 2 \cosh \left(\beta +\sqrt{3} x\right)+\cosh \left[2 \left(\beta +\sqrt{3}
	x\right)\right]  }{ \left[\cosh \left(\beta +\sqrt{3}
	x\right)+2\right] \sinh\left(\beta +\sqrt{3} x\right)},
\end{equation}
which is sufficient to define the corresponding intertwining operators $L_{\pm}$ in (\ref{LW}). Using next the defining relation for the second Hamiltonian $H_2$ we compute the associated second potential from (\ref{H2}) to
\begin{equation}
	V_2 = 3-\frac{3}{2} \text{sech}^2\left(\frac{\beta }{2}+\frac{\sqrt{3} x}{2}\right).
\end{equation}
When shifting the overall energy by $3$, this is an inverse $\cosh^2$-potential discussed in the literature. It possess bound states as treated in section 23 in problem 5 \cite{Landau} and scattering states for which the potential is entirely reflectionless \cite{Lekner}, see also \cite{fluegge}. From \cite{Landau} we read off the eigenfunction for the energy $E$ with potential $U=-U_0 \sech^2 (\alpha x) $, $\alpha, U_0 \in \mathbb{R}$ as
\begin{equation}
\Psi(x) = \left[1-\tanh ^2(\alpha  x)\right]^{\epsilon /2} \, _2F_1\left[\epsilon -s,s+\epsilon +1;\epsilon
+1;\frac{1}{1+e^{2 x \alpha }}\right],
\end{equation}
where $\, _2F_1(a,b;c;z)$ denotes the hypergeometric function and
\begin{equation}
\epsilon :=\frac{ \sqrt{-2 E m}}{\alpha  \hbar }, \qquad  s:= \frac{1}{2} \left(\sqrt{1+\frac{8 m U_0}{\alpha ^2 \hbar ^2}}-1\right) .
\end{equation}
The quantization condition for the energy emerges from the requirement $\lim_{x \rightarrow \pm \infty}\Psi(x) =0$ as $\epsilon-s= -n$ for $n \in \mathbb{N}_0$, which when solved yields
\begin{equation}
	E_n = -\frac{\alpha ^2 \hbar ^2 }{8 m} \left(\sqrt{\frac{8 m U_0}{\alpha ^2 \hbar ^2}+1}-2 n-1\right)^2 .
\end{equation}
As argued in \cite{Landau}, the number of bound states is limited by the constraint $n<s$. When shifting the overall energy by $3$ we obtain $V_2=U$ for the parameter identifications $\hbar=1$, $m=1/2$, $U_0=3/2$, $\alpha=\sqrt{3}/2$. Since for these values $s=1$, there is only one bound state for $n=0$. Thus we obtain the eigensystem solutions
\begin{eqnarray}
	\Phi_0^{(2)}(x) &=& N_0 \text{sech}\left[\frac{1}{2} \left(\beta +\sqrt{3} x\right)\right],  \qquad  E_0^{(2)}=\frac{9}{4}, \\  
	\Phi_1^{(1)}(x)&=& L_+ \Phi_0^{(2)} = N_0 \frac{3 \sqrt{3} \cosh \left[\frac{1}{2} \left(\beta +\sqrt{3} x\right)\right] \coth \left(\beta +\sqrt{3}
		x\right)}{\cosh \left(\beta +\sqrt{3} x\right)+2} ,      \qquad E_1^{(1)}=\frac{9}{4}. \qquad \,\,
\end{eqnarray}
Notice that the singularity in $\Phi_1^{(1)}(x)$ for $x_0 = -\beta \sqrt{3}$ when $\beta \in \mathbb{R}$, inherited form the superpotential $W(x)$, is eliminated in the {$\cal{PT}$}-symmetric solution when $\beta \in i \mathbb{R}$. Since we have obtained a well-defined asymptotically vanishing solution, $\lim_{x \rightarrow \pm \infty} \Phi_1^{(1)}(x) = 0 $ with real eigenvalues, we conclude that the complex type I soliton solution remains stable when linearly perturbed by the shape mode $\Phi_1^{(1)}$. Let us next consider the case when $\lambda$ is taken to be a continuous parameter. 

\subsubsection{Scattering state solutions for type I complex one-soliton solutions}
The scattering states for the inverse $\cosh$-squared potential are discussed in detail in \cite{Lekner}. When adjusting the parameters therein to our equation (\ref{BDtisch}) for $V_2$, we obtain the two scattering solutions
\begin{eqnarray}
	\Psi_e^{(2)}&=& 	\cosh ^2\left(\frac{\hat{x} }{2} \right) \, _2F_1\left[\gamma_-,\gamma_+;\frac{1}{2};-\zeta^2(\hat{x}) \right] 
	     = \cos \left(\tilde{x} \right)-\frac{\sqrt{3} }{2 \kappa} \tanh \left(\frac{\hat{x}}{2}\right) \sin \left( \tilde{x} \right), \qquad \\
	\Psi_o^{(2)}&=& \zeta(\hat{x}) \cosh ^2\left(\frac{\hat{x} }{2}\right) \, _2F_1\left[\frac{1}{2}+\gamma_-,\frac{1}{2}+\gamma_+;\frac{3}{2};-\zeta^2(\hat{x}) \right] \\
	&=& \frac{2 \sqrt{3} \kappa \sin \left( \tilde{x} \right)+3 \tanh \left(\frac{\hat{x}}{2}\right) \cos \left( \tilde{x} \right)}{4 \kappa^2+3} , \notag
\end{eqnarray}
with $\gamma_\pm := 1 \pm i \kappa/\sqrt{3} $, $\zeta(x):=\sinh \left( x/2 \right)$, $\tilde{x}:= \kappa \left(\frac{\beta }{\sqrt{3}}+x\right) $, $\hat{x}:= \beta + \sqrt{3} x$ and continuous eigenvalues $\lambda^2 = 3+ \kappa^2$ for $\kappa$ being a free parameter. As we expect from Abel's identity, the Wronskian for these two solutions is constant, $W(\Psi_e^{(2)},\Psi_o^{(2)})=\Psi_e^{(2)} d\Psi_o^{(2)}/dx -d\Psi_e^{(2)}/dx \Psi_o^{(2)} = \sqrt{3}/2 $, indicating that they are indeed linearly independent. Thus according to (\ref{Super}), the two scattering states for the potential $V_1$ are obtained as
\begin{eqnarray}
	\Psi_e^{(1)}&=& L_+ \Psi_e^{(2)} = \frac{1}{\kappa} \left[ f_-(\hat{x}) \cos \left(\tilde{x} \right) - f_+(\hat{x})\sin \left(\tilde{x} \right) \right], \\ 
	\Psi_o^{(1)}  &=&    L_+ \Psi_o^{(2)} = \frac{2 \sqrt{3}}{3+4\kappa^2} \left[ f_+(\hat{x}) \cos \left(\tilde{x} \right) + f_-(\hat{x}) \sin \left(\tilde{x} \right) \right],
\end{eqnarray}
where
 \begin{equation}
f_+(x):=\frac{3}{2}-\kappa ^2-\frac{9}{2 \cosh \left(x \right)+4}, \qquad  f_-(x):= \frac{3 \sqrt{3} \kappa \cosh (x) \coth \left(\frac{x}{2}\right)}{2 \cosh (x)+4} .
 \end{equation}
 Once again the singularity at $x_0 = -\beta \sqrt{3}$ is eliminated in the {$\cal{PT}$}-symmetric solution by taking $\beta \in i \mathbb{R}$.
Now the Wronskian for these two solutions is computed to $W(\Psi_e^{(1)},\Psi_o^{(1)})=(3 + \kappa^2)\sqrt{3}/2  $. When $\kappa = \kappa_0:=\pm i \sqrt{3}$ we recover the zero mode $\Phi_0^+$ from $\Psi_e^{(1)}$, whereas $ \lim_{x \rightarrow \kappa_0 } \Psi_o^{(1)}(x)=0$.

 Since the scattering solutions are well-defined with real continuous eigenvalues one may be tempted to conclude that the complex type I soliton solution remains stable when linearly perturbed by either of the two scattering solutions  $\Psi_e^{(1)}$ or $\Psi_o^{(1)}$. However, as the limits for the asymptotic values of $x$ are not well defined, the energies of the perturbed solution (\ref{enn}) do not give definite values so that the perturbation by these scattering solutions leads to ill-defined objects. A natural way to overcome this issue would be to restrict the domain of the theory to finite interval. 
 
 An interesting solution arises for $\kappa=0$
 \begin{equation}
 	\lim_{\kappa \rightarrow 0} \Psi_o^{(1)} = \sqrt{3} \left(1-\frac{3}{\cosh \left(\beta +\sqrt{3} x\right)+2}\right) =: \Psi_0^{(1)} ,
 \end{equation}
which has a well-defined finite limit $\sqrt{3}$ for $x \rightarrow \pm \infty$. Since the derivative of this solution is asymptotically vanishing as well as the derivative of the static solution $\phi_I$, the second line in (\ref{enn}) is vanishing up to order $\varepsilon^2$ so that the energy of the perturbed solution is the same as the original solution. Since we have also a positive eigenvalue for the auxiliary eigenvalue problem we conclude that the solution remains stable when perturbed by $\Psi_0^{(1)}$.

\subsection{Complex one-soliton solutions of type II} 

\begin{figure}[h]
	\centering         
	\begin{minipage}[b]{0.8\textwidth}           
		\includegraphics[width=\textwidth]{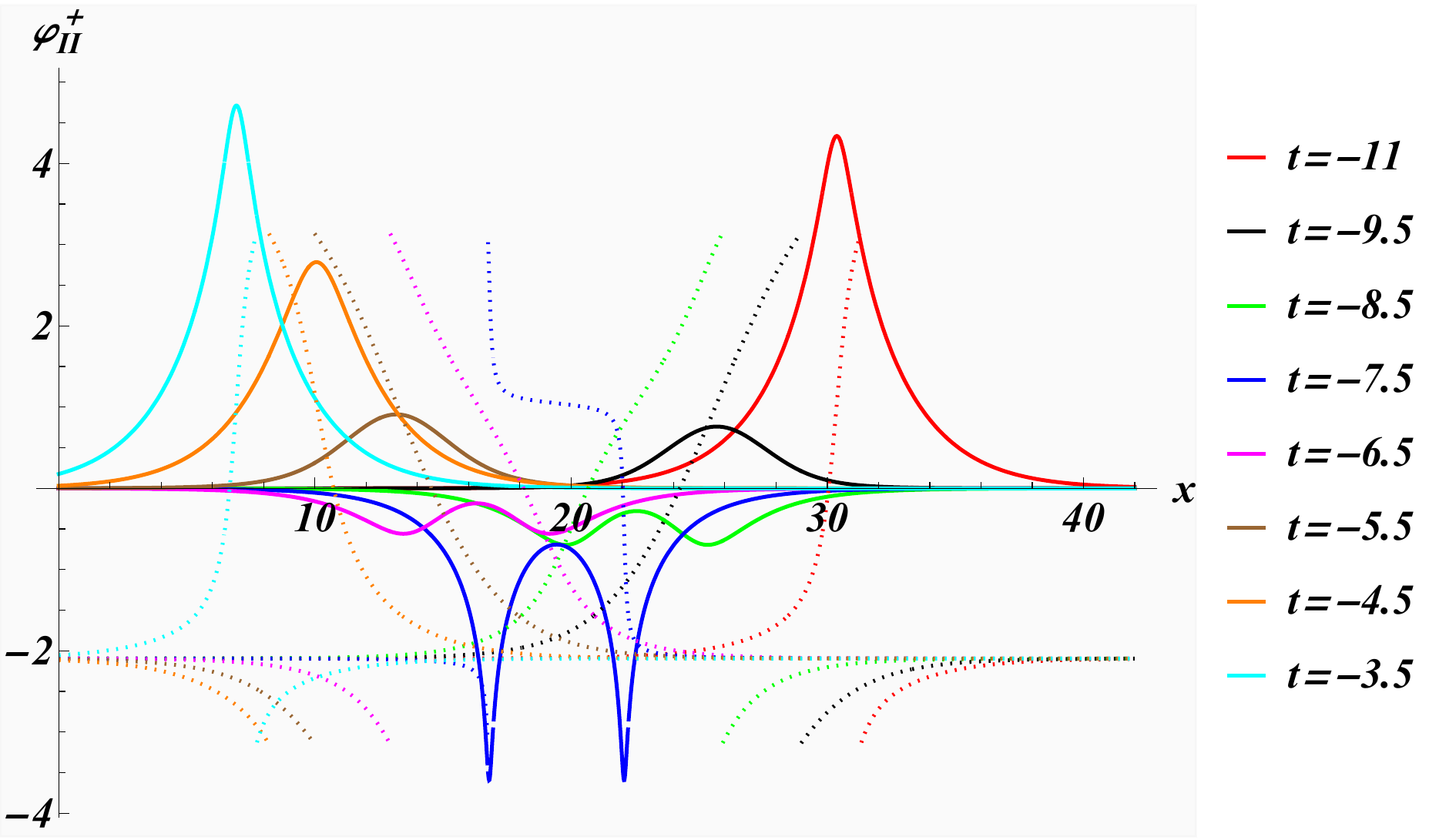}
	\end{minipage}    
	\caption{Complex one-soliton solutions of type II (\ref{sol2}) with $\beta=2$, $k=1/2$ for the Bullough-Dodd classical equation of motion (\ref{BDequ}). Solid lines correspond to real and dotted lines to imaginary parts. }
	\label{BDsolution3}
\end{figure}

The second type of one-soliton solutions obtained from (\ref{Ansatz}) by fixing the unknown constants from direct substitution into the Bullough-Dodd equation (\ref{BDequ}) are always complex and do not possess a real regime in their range
\begin{equation}
	\varphi_{II}^{\pm} (x,t)	= \ln \left\{ \omega^{\mp 2} \left[1-\frac{6   e^{t \sqrt{k^2+ 3 \omega^{\pm 1}}+k
			x \pm \beta}}{\left(1+   e^{t \sqrt{k^2+3 \omega^{\pm 1}}+k x \pm \beta}\right)^2}\right]\right\} . \label{sol2}
\end{equation}
Here $\omega=e^{i \pi/3}$ denotes the third root of unity. A sample solution is depicted in figure \ref{BDsolution3}, from which we observe that the real part of this solution is an oscillation between a regular shaped soliton solution and a double peakon solution. 

The energies when directly computed with (\ref{ener}) for the Lagrangian densities $\mathcal{L}_{\text{BD}}$ are infinite. However, when shifting the vacuum from $\varphi_0=0$ to the complex plane  $\varphi_0^{\pm}= \mp i 2 \pi /3$ amounts to defining the new Lagrangians densities
\begin{equation}
	\mathcal{L}_{\text{BD}}^{\pm} = \mathcal{L}_{\text{BD}} + \frac{3}{2} \left( \omega^{\mp 2} - 1 \right) .
\end{equation}
Computing the energies of the solution $\varphi_{II}^{\pm}$ with these Lagrangians leads to the same real values as those computed for the type I solution with $\mathcal{L}_{\text{BD}}$
\begin{equation}
	E^{+} [ \varphi_{II}^{+} ] = 	E^{-} [ \varphi_{II}^{-} ] = - 6 |k| . \label{energII}
\end{equation}
The ${\cal{PT}}$-symmetry for the individual Lagrangians is now broken, but instead we have ${\cal{PT}}$: $\mathcal{L}_{\text{BD}}^{+} \rightarrow \mathcal{L}_{\text{BD}}^{-}$. Since the energies are real this suggests that the criteria provided in \cite{CFTSkyrmions} may still be slightly enlarged. One may easily convince oneself that the energies of two ${\cal{PT}}$-symmetrically related degenerate solutions are real even when they originate from two different ${\cal{PT}}$-symmetrically related Hamiltonians. Having obtained real energies suggests that these solutions are potentially well-defined physical objects, but let us investigate whether the solutions are also stable.

\subsubsection{Zero modes for complex type II one-soliton solutions}
The static solutions corresponding to $\varphi_{II}^{\pm} (x,t)$ are now obtained as 
\begin{equation}
	\phi_{II}^{\pm} (x)	= \ln \left[ \omega^{\mp2}  \frac{\cos(\sqrt{ 3 }\omega^{\pm 1/2} x\pm  \beta ) -2}{ \cos( \sqrt{ 3 }\omega^{\pm 1/2} x\pm \beta ) +1}       \right] . \label{solII}
\end{equation}
Taking $\beta \in \mathbb{R}$ the two solutions are related to each other by the ${\cal{PT}}$-symmetry in (\ref{PT}) as $ {\cal{PT}}: \phi_{II}^{+} \rightarrow \phi_{II}^{-}$. Alternatively, we may also change the sign in $\beta$ in $\phi_{II}^{-}$ and take $\beta \in i \mathbb{R}$ in which case the two solutions are still related by the same symmetry. The potentials for the eigenvalue equation as introduced in (\ref{BDtisch}) result to
\begin{equation}
V_1^{\pm} = \omega^{\mp 2} \left\{ 1-\frac{3}{ \cos\left(\sqrt{ 3 }\omega^{\pm 1/2}x \pm \beta\right)+ 1}+
 2 \frac{ [\cos(\sqrt{ 3 }\omega^{\pm 1/2}x \pm \beta )+1]^2}{[\cos(\sqrt{ 3 }\omega^{\pm 1/2}x \pm \beta )-2]^2}  \right\} .
\end{equation}
Since the two potentials are not ${\cal{PT}}$-symmetric by themselves one may be tempted to conclude already at this stage that there are no real eigenvalue solutions to (\ref{BDtisch}). However, all we can deduce with certainty at this point is that the symmetry as defined in (\ref{PT}) is broken and will not ensure the reality of $\lambda$. We can not guarantee that there are no other types antilinear symmetries that would explain a real spectrum. 

Thus let us show explicitly how the eigensystem may be solved in this case.  The function $\psi$ in (\ref{zeromodecal}) is now taken to $\psi^{\pm}(x,a)=\phi^{\pm}(x+a \omega^{\mp 1/2})$, from which we easily calculate the zero mode to
\begin{equation}
	\Phi_0^{\pm}(x) = \mp \frac{3^{3/2} \tan \left[\frac{1}{2} (\sqrt{ 3 }\omega^{\pm 1/2}x \pm \beta  )\right]}{  \cos \left(\sqrt{ 3 }\omega^{\pm 1/2}x \pm \beta \right)-2} . \label{phizero2}
\end{equation}
We notice that in these solutions there is no need to take $\beta$ to be purely imaginary as the singularities in the zero mode are already regularized because $x$ is multiplied by a complex constant.

\subsubsection{Bound state solutions for type II complex one-soliton solutions}
Next we employ relation (\ref{LW}) to compute the superpotentials from the zero modes
\begin{equation}
	W^{\pm}(x) = \pm \frac{\sqrt{ 3 }\omega^{\pm 1/2}}{2} \frac{ [ 3+\cos \left[2 \left(\sqrt{ 3 }\omega^{\pm 1/2}x \pm \beta \right) \right]-2 \cos \left( \sqrt{ 3 }\omega^{\pm 1/2}x \pm \beta\right) ] }{ [\cos\left( \sqrt{ 3 }\omega^{\pm 1/2}x \pm \beta\right)-2] \sin \left(\sqrt{ 3 }\omega^{\pm 1/2}x \pm \beta\right)},
\end{equation}
from which we calculate the corresponding intertwining operators $L_{\pm}$ in (\ref{LW}). The partner potentials are then obtained by a direct calculation from (\ref{H2}) to
\begin{equation}
	V_2^{\pm} =  \frac{3 \omega ^{\mp 2}}{ 1-\sec \left( \sqrt{ 3 }\omega^{\pm 1/2} x \pm \beta \right) }=\frac{3}{2} \frac{\omega^{\pm 1}}{ \sin ^2\left[\frac{1}{2} (\pm \beta +\sqrt{ 3 }\omega^{\pm 1/2} x) \right]}-3 \omega^{\pm 1} .
\end{equation}
Again, this is a well studied potential corresponding to a special version of the P\"oschl-Teller potential \cite{PTe,fluegge,correa2012spectral,dey2013bohmian}. Adapting the shift and scaling to the solutions that can be found in the literature, we obtain the eigensystem solutions
\begin{eqnarray}
	\Phi_0^{(2) \pm } &=& N_0 \frac{1}{ \sin \left[ \frac{1}{2} \left(\pm  \beta +\sqrt{3} \omega^{\pm 1/2} x \right)\right]} ,  \qquad  E_0^{(2)}=-\frac{9}{4} \omega^{\pm 1}, \\  
	\Phi_1^{(1) \pm}&=& L_+ \Phi_0^{(2)} =  \frac{N_0}{2} \frac{3^{3/2}  \omega^{\pm1/2}  \cos \left(\pm \beta +\sqrt{3}  \omega^{\pm 1/2} x \right) }{ \left[2-\cos \left(\pm \beta +\sqrt{3} \omega^{\pm 1/2} x \right)\right] \sin \left[\frac{1}{2} \left(\pm \beta
		+\sqrt{3} \omega^{\pm 1/2} x \right)\right]  } ,    \,\,   E_1^{(1)}=E_0^{(2)}.\notag
\end{eqnarray}
The solutions are asymptotically vanishing, $\lim_{x \rightarrow \pm \infty} \Phi_1^{(1) \pm}(x) = 0 $ and come in complex conjugate pairs, which is the signature property of a spontaneously broken ${\cal{PT}}$-symmetry. Instead we have ${\cal{PT}}: \Phi_1^{(1) +}(x) \rightarrow \Phi_1^{(1) -}(x)$. We conclude that the complex soliton solutions of type II become unstable when perturbed by the bound state solutions $\Phi_1^{(1) \pm}$.   
\subsubsection{Scattering solutions for type II complex one-soliton solutions}
Finally we discuss the scattering solutions for the auxiliary Sturm-Liouville eigenvalue equation obtained from the type II complex one-soliton solutions. By similar means as in the previous section we exploit the fact that the potential becomes a shifted and scaled inverse $\sin$-squared potential. For the $V_1^{\pm}$-potential we then find the two linearly independent scattering solutions 
\begin{eqnarray}
	\Psi_e^{(2)\pm}&=&  	g^{\pm}(x)  \sin (\kappa x)   +  h^{\pm}(x)  \cos (\kappa x),  \\
	\Psi_o^{(2) \pm}&=&  	g^{\pm}(x)  \cos (\kappa x)   +  h^{\pm}(x)  \sin (\kappa x), 
\end{eqnarray}
with
\begin{eqnarray}
	g^{\pm}(x) &=& \pm \frac{3^{3/2} \omega^{\pm 1/2}  \kappa}{2} \frac{ \cos \left(\beta + \sqrt{3} \omega^{\pm 1/2}  x  \right) \tan \left[\frac{1}{2} \left( \beta + \sqrt{3} \omega^{\pm 1/2} x \right)\right]}{ \cos \left( \beta + \sqrt{3} \omega^{\pm 1/2} x \right) -2 }   \\  
		h^{\pm}(x)&=& \frac{3}{2} \omega^{\pm 1} \left[ \frac{3}{2- \cos \left( \beta + \sqrt{3} \omega^{\pm 1/2} x  \right)} -1    \right] -\kappa^2
\end{eqnarray}
for the complex continuous eigenvalues $(\lambda^\pm)^2 =\kappa^2 - 3 \omega^\pm $. One may be tempted to choose the free parameter $\kappa$ in such a way that $(\lambda^\pm)^2$ becomes real. However, when $\kappa$ acquired a complex part the solutions are easily seen to diverge at $x \rightarrow \pm \infty$. Thus overall we conclude that the complex type II one-solitons solutions become unstable when linearly perturbed with these scattering solutions $\Psi_{e,o}^{(2)\pm}$.  

\section{Conclusions}
We investigated several types of complex one-soliton solutions in the Bullough-Dodd model with regard to their stability when linearly perturbed. The type I solutions were previously known, whereas the type II solutions have been newly constructed. The overall energy for both types of solutions were found to be real when shifting the vacuum of the potential appropriately. 

We found that the auxiliary Sturm-Liouville eigenvalue equation of the stability analysis for the type I solutions can be supersymmetrically mapped to an isospectral partner system involving a shifted and scaled inverse $\cosh$-squared potential. Remarkably these potentials also emerge in the stability analysis of the $\phi^4$-theory and sine-Gordon model, but simply with different scalings and overall shifts \cite{jackiw1977q}. We concluded that the type I solutions remain stable when linearly perturbed with bound state and a scattering solutions with vanishing continuous parameter.

For the type II solutions the original auxiliary Sturm-Liouville eigenvalue equation can be supersymmetrically mapped to a complex shifted and scaled inverse $\sin$-squared potential. The corresponding bound state and scattering solutions for these equations have complex eigenvalues or are asymptotically divergent, such when used to perturb the type II solutions they become unstable or the energies become infinite. 

The underlying ${\cal{PT}}$-symmetry serves for several purposes: i) When introducing complex rather than real shifts the singularities in the solutions are regularized. ii) As previously pointed out, it governs the reality of the energy of the solutions, but as we found here the set criteria provided in \cite{CFTSkyrmions} may still be enlarged. The energies of two ${\cal{PT}}$-symmetrically related degenerate solutions are guaranteed to be real even when they solve the Euler-Lagrange equations of two ${\cal{PT}}$-symmetrically related Hamiltonians that might even be different. iii) As we have seen here the ${\cal{PT}}$-symmetry also governs the stability of the solutions. However, in order to obtain stable solutions, that is real eigenvalues in the auxiliary Sturm-Liouville equation, the perturbation must be ${\cal{PT}}$-symmetric by themselves. 

Here we observed that the ${\cal{PT}}$-symmetry of the soliton solution $\varphi$ was inherited by the perturbing field $\chi$. The type I solutions and their perturbing fields were found to be ${\cal{PT}}$-symmetry by themselves, which guaranteed the reality of the energy and their stability. On the other hand the type II solutions were mapped into different degenerate solutions, so that the reality of their energy was still ensured. However, on the level auxiliary of the Sturm-Liouville equation the ${\cal{PT}}$ is broken for a specific solution so that the eigenvalues become complex and the solutions unstable. 

\medskip

\noindent \textbf{Acknowledgments:} FC was partially supported by Fondecyt
grant 1211356. TT is supported by EPSRC grant EP/W522351/1. 

\newif\ifabfull\abfulltrue

\end{document}